\documentclass[reprint, amsmath,amssymb, aps, prl, superscriptaddress]{revtex4-2}

\usepackage{graphicx}
\usepackage{dcolumn}
\usepackage{bm}
\usepackage{multirow}
\usepackage{braket}
\usepackage{amsfonts}
\usepackage{amssymb}
\usepackage{amsthm}
\usepackage{array}
\usepackage{amsmath}
\usepackage{verbatim}
\usepackage{color}

\newcommand{\affa}{State Key Laboratory of Low Dimensional Quantum Physics, Department of Physics, Tsinghua University, Beijing 100084, China}
\newcommand{\affb}{Beijing Academy of Quantum Information Sciences, Beijing 100193, China}
\newcommand{\affc}{Hefei National Laboratory, Hefei 230088, P. R. China}
\newcommand{\affd}{Frontier Science Center for Quantum Information, Beijing 100084, China}
\newcommand{\affe}{Department of Physics, Renmin University, Beijing 100872, China}
\newcommand{\affg}{MOE Key Laboratory of Fundamental Physical Quantities Measurement, Hubei Key Laboratory of Gravitation and Quantum Physics, PGMF, Institute for
Quantum Science and Engineering, School of Physics, Huazhong University of Science and Technology,Wuhan 430074, China}
\newcommand{\affh}{Shenzhen Institute for Quantum Science and Engineering, Southern University of Science and Technology, Shenzhen 518055, China}
\newcommand{\affi}{Graduate School of China Academy of Engineering Physics, Beijing 100193, China}

\def\bra#1{\langle #1|}
\def\ket#1{\left|#1 \right>}

\begin{document}

\title{Error-Mitigated Quantum Simulation of Interacting Fermions with Trapped Ions}


\author{Wentao Chen}
\thanks{These two authors contributed equally to the work.}
\address{\affa}
\author{Shuaining Zhang}
\thanks{These two authors contributed equally to the work.}
\email{zhangshuaining@ruc.edu.cn}
\address{\affe}
\address{\affa}
\address{\affb}
\author{Jialiang Zhang}
\address{\affa}
\author{Xiaolu Su}
\address{\affa}
\author{Yao Lu}
\address{\affh} \address{\affa}
\author{Kuan Zhang}
\address{\affg} \address{\affa}
\author{Mu Qiao}
\address{\affa}
\author{Ying Li}
\email{yli@gscaep.ac.cn }
\address{\affi}
\author{Jing-Ning Zhang}
\email{zhangjn@baqis.ac.cn}
\address{\affb}
\author{Kihwan Kim}
\email{kimkihwan@mail.tsinghua.edu.cn}
\address{\affa}
\address{\affb}
\address{\affc}
\address{\affd}

\begin{abstract}
Quantum error mitigation has been extensively explored to increase the accuracy of the quantum circuits in noisy-intermediate-scale-quantum (NISQ) computation, where quantum error correction requiring additional quantum resources is not adopted. Among various error-mitigation schemes, probabilistic error cancellation (PEC) has been proposed as a general and systematic protocol that can be applied to numerous hardware platforms and quantum algorithms. However, PEC has only been tested in two-qubit systems and a superconducting multi-qubit system by learning a sparse error model. Here, we benchmark PEC using up to four trapped-ion qubits. For the benchmark, we simulate the dynamics of interacting fermions with or without spins by applying multiple Trotter steps. By tomographically reconstructing the error model and incorporating other mitigation methods such as positive probability and symmetry constraints, we are able to increase the fidelity of simulation and faithfully observe the dynamics of the Fermi-Hubbard model, including the different behavior of charge and spin of fermions. Our demonstrations can be an essential step for further extending systematic error-mitigation schemes toward practical quantum advantages. 
\end{abstract}

\maketitle

\section{Introduction} 
Recently, quantum computation in the noisy intermediate scale quantum (NISQ) regime has been actively developed for the possibility of reaching practical quantum advantages without quantum error correction \cite{preskill2018quantum,bharti2022noisy}. In this direction, quantum advantages with random gate sampling algorithms in superconducting systems have been demonstrated \cite{arute2019quantum,wu2021strong}. The possibility of surpassing conventional classical computation in solving useful and practical problems has been seriously explored \cite{daley2022practical}. In particular, a quantum-classical hybrid method that combines the advantages of classical and quantum computation, such as the variational quantum algorithm (VQA), has been theoretically and experimentally explored to reach higher performance than with only classical computers \cite{peruzzo2014variational,shen2017quantum,cerezo2021variational,tilly2022variational}. One of the critical issues in NISQ computation is to obtain results with high precision from quantum devices without quantum error correction \cite{shor1995scheme,steane1996error}. 

Various error-mitigation methods without using additional quantum resources for quantum error corrections have been proposed and applied to improve the computation results of quantum devices \cite{temme2017error,li2017efficient,endo2018practical,cai2022quantum}. Representative error-mitigation methods include zero-noise extrapolation \cite{temme2017error,li2017efficient,endo2018practical,cai2022quantum,kandala2019error,kim2021scalable}, probabilistic error cancellation (PEC) \cite{temme2017error,li2017efficient,endo2018practical,cai2022quantum,song2019quantum,zhang2020error,piveteau2022quasiprobability,takagi2021fundamental}, virtual distillation \cite{huggins2021virtual}, etc. In particular, PEC is considered the general and systematic method for mitigating errors, which are not dependent on details of the physical implementation as long as errors are properly characterized. PEC consists of characterizing errors of each quantum gate, decomposing the inverse of the errors into combinations of basis operations, and sampling the operations with the characterized weights. In this way, we can estimate the expectation values of the desired observables without errors. PEC has been theoretically analyzed in detail and shown to be optimal for dephasing dominant noise situations \cite{temme2017error,li2017efficient,piveteau2022quasiprobability,takagi2021fundamental}. However, it has been experimentally tested with only two-qubit systems of superconducting circuits and trapped ions  \cite{song2019quantum,zhang2020error} and a 
multi-qubit system of the superconducting circuit via fitting a sparse error model motivated by the processor topology \cite{berg2022probabilistic}. 

Systematic error-mitigation schemes like PEC would enhance the performance of quantum computation in the NISQ regime drastically. Many important quantum algorithms have been developed based on the fact that quantum dynamics can be efficiently simulated in quantum computation, which also inspires VQA designs \cite{peruzzo2014variational,shen2017quantum,cerezo2021variational,tilly2022variational}. In digital quantum simulation, the time evolution of a certain Hamiltonian is typically implemented by the Trotter-Suzuki expansion~\cite{suzuki1990fractal}. Besides the errors in the expansion, imperfect primitive quantum gates also degrade the fidelity of the quantum simulation and eventually lead to faulty results. Due to this reason, it is an essential and urgent task to design and benchmark error-mitigation techniques that are both systematic and efficient in the NISQ era. As an instance for benchmark, the Fermi-Hubbard model \cite{hubbard1963electron} is particularly interesting. This model is one of the fundamental models for interacting electrons, and it has been extensively studied in classical manners, either analytically~\cite{lieb1968absence,metzner1989correlated,muller1989correlated} or numerically~\cite{leblanc2015solutions}. As experimental quantum technologies develop, understanding and insights for strongly-correlated quantum systems and high-temperature superconductivity~\cite{lee2006doping} can be explored through quantum simulation of the Fermi-Hubbard model eventually beyond the classical computational capability. This model has been utilized in experimental demonstration of various quantum platforms~\cite{casanova2012quantum,barends2015digital,mazurenko2017cold,hensgens2017quantum}. Recently, a combination of non-PEC methods has been applied to enhance simulation performance and observe the dynamics of the Fermi-Hubbard model in a superconducting system \cite{arute2020observation}. 

\begin{figure*}[htp!]
\centerline{\includegraphics[width=\textwidth]{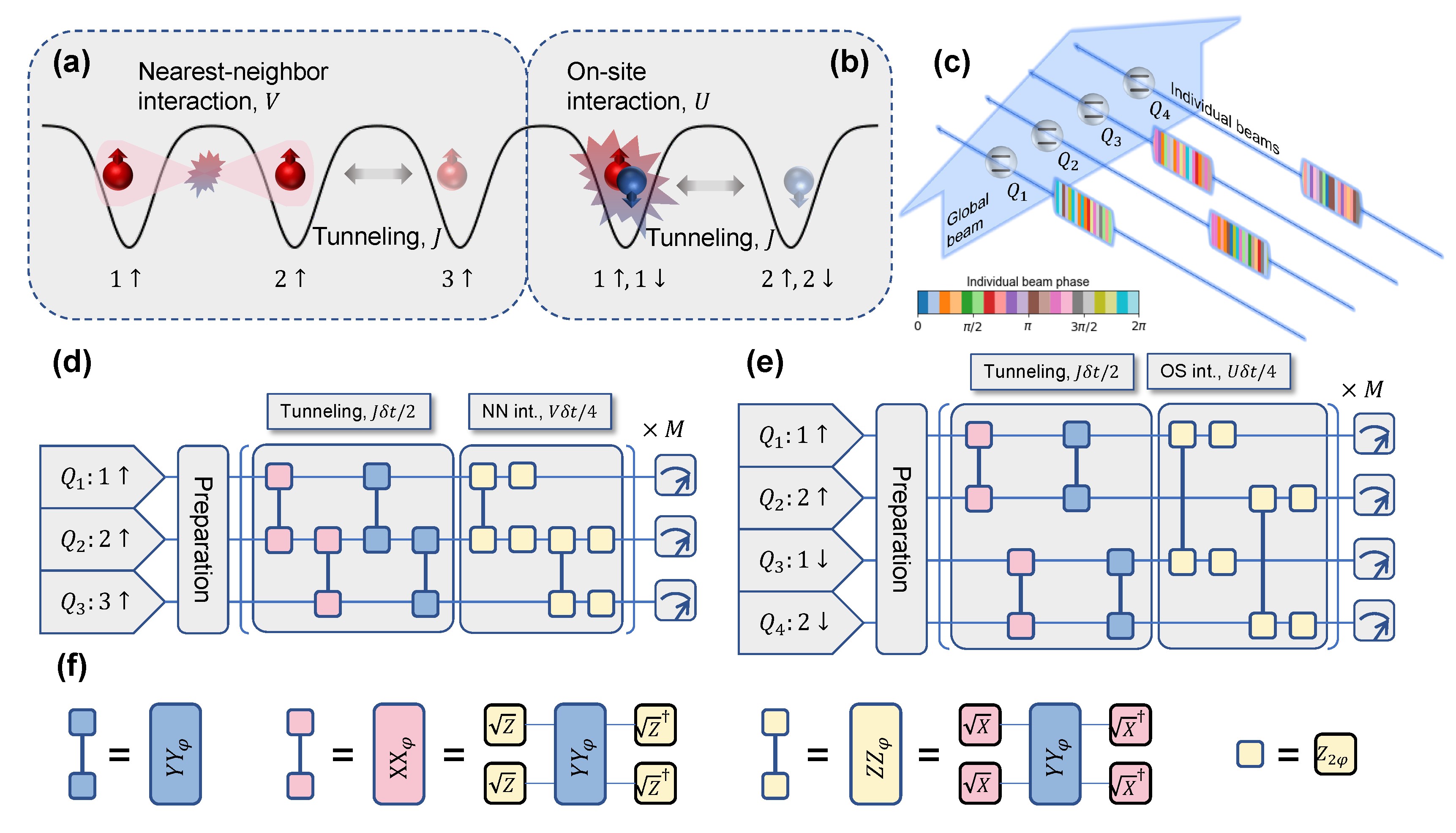}}
\caption{(a) Spinless and (b) spin-$\frac{1}{2}$ fermions moving in a one-dimensional lattice. The system dynamics can be unified by the extended Fermi-Hubbard model in Eq.~(\ref{eq:1}), with the tunneling strength denoted by $J$ and the on-site and the nearest-neighbor interaction strengths by $U$ and $V$, respectively. (c) Trapped-ion system with four ion-qubits. The ions form a 1-D crystal and each ion represents a qubit. The ion qubits can be manipulated by Raman laser beams, consisting of a global beam and individual beams, with the former covering the whole ion crystal and each of the latter focusing on a single ion. M{\o}lmer-S{\o}rensen gate $YY_\varphi(m, n)\equiv \exp\left(i\varphi\hat\sigma^y_m\otimes\hat\sigma^y_n\right)$ can be implemented between any qubit pair $(Q_m, Q_n)$ by modulating the phase of the individual beams. (d) Quantum circuit for simulating the dynamics of a three-site single-component extended Fermi-Hubbard model. With the Trotter-Suzuki expansion, the evolution is approximately implemented by $M$ identical steps. (e) Quantum circuit for a two-site double-component Fermi-Hubbard model. (f) Quantum circuit compilation. While the two-qubit entangling gate provided by the native gate set of the trapped-ion system is $YY_\varphi$, $XX_\varphi$ and $ZZ_\varphi$ can be constructed by $YY_\varphi$ and appropriate single-qubit rotations, such as $\sqrt{Z}\equiv\exp\left(-i\frac{\pi}{4}\hat\sigma_z\right)$ and $\sqrt{X}\equiv\exp\left(-i\frac{\pi}{4}\hat\sigma_x\right)$. The single-qubit gates (yellow squares) in (d, e) are the $\hat\sigma_z$-rotations, i.e. $Z_{2\varphi}\equiv\exp\left(-i\varphi\hat\sigma_z\right)$.}
\label{fig:device_schematic}
\end{figure*}

In this paper, we experimentally study the possibility and limitations of PEC by applying the method to a four-qubit system with trapped ions \cite{lanyon2011universal,lu2019global}. In particular, we focus on simulating the dynamics of the 1D Fermi-Hubbard model by using multiple Trotter steps. We find the PEC method with a tomographically reconstructed error model significantly improves the fidelity of the simulation with up to three ion qubits. However, with four qubits, the improvement of fidelity by the PEC method alone is much worse than those for two- and three-qubit systems. This is because the tomographic characterization of the noisy gates is not precise enough to capture the time correlation and the cross-talk errors. With the application of total-spin and particle-number conservation \cite{bonet2018low,mcardle2019error,cai2022quantum} as well as positivity constraint, however, the enhanced PEC method provides recovery for the ideal dynamics of the 1D spin-Fermi-Hubbard model, where the spin and the charge behave differently. 

\section{Fermi-Hubbard Model and Experimental Realization with Trapped ions}
Here we consider the extended Fermi-Hubbard (eFH) model, which describes a system of fermions moving in a one-dimensional lattice, as illustrated in Fig.~\ref{fig:device_schematic}a and Fig.~\ref{fig:device_schematic}b for one-component and two-component fermions. The one-component fermions can be considered spinless ones.  
The Hamiltonian reads
\begin{eqnarray}
\hat H_{\rm eFH}&=&-J\sum_{l,\lambda}\left(\hat c_{l,\lambda}^\dag\hat c_{l+1,\lambda}+{\rm h.c.}\right)\nonumber\\
&&+V\sum_{l,\lambda,\lambda'}\hat n_{l,\lambda}\hat n_{l+1,\lambda'}+U\sum_l\hat n_{l,\uparrow}\hat n_{l,\downarrow},
\label{eq:1}
\end{eqnarray}
with $\hat n_{l,\lambda}\equiv\hat c_{l,\lambda}^\dag\hat c_{l,\lambda}$, where $\hat c_{l,\lambda}$ ($\hat c_{l,\lambda}^\dag$) annihilates (creates) a fermion with spin $\lambda$ ($\lambda=\uparrow,\downarrow$) on the $l$-th site. Here $J$ is the nearest-neighbor tunneling strength, and $U$ and $V$ respectively quantify the strength of the on-site and the nearest-neighbor interaction.

We can map the fermion system to a qubit system by using the Jordan-Wigner transformation,
\begin{eqnarray}
\hat c_{l,\uparrow}&=&\prod_{n=1}^{l-1}\hat\sigma_n^z\hat\sigma_l^-,\quad \hat n_{l,\uparrow}=\frac{1}{2}\left(1-\hat\sigma_l^z\right),\\
\hat c_{l,\downarrow}&=&\prod_{n=1}^{L+l-1}\hat\sigma_n^z\hat\sigma_{L+l}^-,\quad\hat n_{l,\downarrow}=\frac{1}{2}\left(1-\hat\sigma_{L+l}^z\right),\nonumber
\end{eqnarray}
with $\hat\sigma_{n}^\pm=\left(\hat\sigma_n^x\pm i\hat\sigma_n^y\right)/2$ and $\hat\sigma_n^{x,y,z}$ being the Pauli operators on the $n$-th qubit. Thus generally speaking, we need $N=2L$ qubits to simulate the dynamics of an $L$-site fermionic chain. 

As shown in Fig.~\ref{fig:device_schematic}c, we use $^{171}{\rm Yb}^+$ ions for qubit systems. The $^{171}{\rm Yb}^+$ ions are confined in a linear Paul trap. The hyperfine levels in the $^2S_{1/2}$ ground-state manifold, i.e. $\ket{F=0,m_F=0}$ and $\ket{F=1,m_F=0}$, are encoded as the qubit $\left\{\ket{0},\ket{1}\right\}$, which can be initialized to $\ket{0}$ by optical pumping and disgustingly detected by fluorescence measurement via multi-channel photo-multiplier tube \cite{olmschenk2007manipulation}.

The qubit-system Hamiltonian becomes
\begin{eqnarray} \label{eq:3}
\hat H_{\rm qb}&=&\hat H_X+\hat H_Y+\hat H_Z,\\
\hat H_X&=&\frac{J}{2}\sum_{l=1}^{L-1}\left(\hat\sigma_l^x\hat\sigma_{l+1}^x+\hat\sigma_{L+l}^x\hat\sigma_{L+l+1}^x\right)\nonumber\\
\hat H_Y&=&\frac{J}{2}\sum_{l=1}^{L-1}\left(\hat\sigma_l^y\hat\sigma_{l+1}^y+\hat\sigma_{L+l}^y\hat\sigma_{L+l+1}^y\right)\nonumber\\
\hat H_Z&=&\frac{V}{4}\sum_{l=1}^{L-1}\left[\left(1-\hat\sigma_l^z\right)\left(1-\hat\sigma_{l+1}^z\right)\right.\nonumber\\
&&\left.+\left(1-\hat\sigma_{L+l}^z\right)\left(1-\hat\sigma_{L+l+1}^z\right)\right]\nonumber\\
&&+\frac{U}{4}\sum_{l=1}^{L}\left(1-\hat\sigma_l^z\right)\left(1-\hat\sigma_{L+l}^z\right).\nonumber
\end{eqnarray}
Note that according to the orientation of the Pauli operators, we divide the qubit-system Hamiltonian $\hat H_{\rm qb}$ into three parts, each of which contains mutually commutative terms.

To extract dynamical properties, the essential part is to simulate the evolution operator $\hat U(t)=\exp\left(-i\hat H_{\rm qb}t\right)$ with a universal gate set available in a certain quantum computational platform. One of the well-known solutions is to use the Trotter-Suzuki decomposition. The first-order Trotter-Suzuki decomposition of $\hat U(t)$ reads 
\begin{eqnarray}
\hat U(t)=\left(e^{-i\hat H_Z\delta t}e^{-i\hat H_Y\delta t}e^{-i\hat H_X\delta t}\right)^M+{\mathcal O}\left(M\delta t^2\right),
\label{eq:4}
\end{eqnarray}
with the Trotter step being $\delta t=t/M$. Figs.~\ref{fig:device_schematic}d and e show examples of the circuits for simulating the dynamics of 3-site spinless and 2-site spin-$1/2$ fermionic systems, with the native gate set including the M{\o}lmer-Sorensen gates, i.e. ${XX}_\varphi$, ${YY}_\varphi$ and ${ZZ}_\varphi$, and single-qubit rotations.  By applying the amplitude-shaped \cite{schindler2008} global and individual laser beams on the desired ions and driving the transverse motional modes \cite{sorensen1999quantum,kim2009entanglement}, we can realize the native  M\o lmer-S\o rensen $YY$-gate (${YY}_\varphi$).  The gate ${XX}_\varphi$ and ${ZZ}_\varphi$ are composed with ${YY}_\varphi$ and corresponding single qubit rotations, as shown in Fig.~\ref{fig:device_schematic}f.

In this experiment, we consider two scenarios, one for spinless fermions, and the other for two-component fermions. In the former case, the on-site interaction strength naturally vanishes, i.e. $U=0$, due to the fermionic anti-commutation relation. In this case, we use $N=L$ ion-qubits to simulate the dynamics of a spinless fermionic chain with $L$ sites. While for the latter case, we only consider the on-site interaction, since the interaction strength decreases as the distance increases and thus $U\gg V$ in most circumstances. 

\section{Probabilistic error cancellation in trapped ion system } 
Here, we introduce the PEC error-mitigation method, which systematically recovers ideal expectation values of target quantum circuits with the experimental noisy gates \cite{endo2018practical}. The essential idea is to first characterize experimental noisy gates, then identify the noise parts by comparing the ideal and the noisy gates, and then decompose the inverse of the noise parts into combinations of basis operations. Then, each of the ideal gates in target quantum circuits is replaced by a corresponding noisy one followed by basis operations sampled with probability distributions obtained from the inverse-noise decomposition. At last, the ideal expectation value can be obtained by adding up those of the sampled circuits with appropriate weights. 

In circuit quantum computation, a computational task can be compiled into a quantum circuit consisting of single- and two-qubit gates. In the trapped ion system, the error of the two-qubit gate is larger than the single-qubit one \cite{zhang2020error}, and we can mitigate only the errors in two-qubit gates. To characterize the experimental noisy gates, we perform quantum process tomography (QPT), under the Pauli error assumption, to obtain the Pauli transfer matrices (PTMs) of the M{\o}lmer-S{\o}rensen $YY$ gates on different qubit pairs, i.e. $R_{YY_{\varphi}^{\rm ns}(m,n)}$ for the qubit pair $(Q_m, Q_n)$. Together with the ideal PTM $R_{YY_\varphi}$, we can formally define an error operator for each of the noisy gates, with the PTM being
\begin{eqnarray}
R_{{\mathcal E}_\varphi(m,n)}=R_{YY_{\varphi}^{\rm ns}(m,n)}R_{YY_{\varphi}}^{-1}.\label{eq:PTM_error}
\end{eqnarray}
Note that under the Pauli-error assumption, both $R_{{\mathcal E}}$ and its inverse $R_{{\mathcal E}}^{-1}$ can be decomposed into linear superpositions of PTMs of two-qubit Pauli operators $P_i \in\{II,IX,IY,...,ZZ\}$. As a result, ideal PTMs can be rewritten in terms of the PTMs of the noisy gate and the corresponding error operators,
\begin{eqnarray}
R_{YY_\varphi}=R_{{\mathcal E}_\varphi(m,n)}^{-1}R_{YY_\varphi^{\rm ns}(m,n)}
\end{eqnarray}
with the inverse-error decomposition
\begin{eqnarray}
R_{{\mathcal E}_\varphi(m,n)}^{-1}=\sum_{i=0}^{15}q_i(m, n)R_{P_i}.
\end{eqnarray}
Here the quasi-probabilities $q_i(m, n)$ are real and satisfy $\sum_i q_i(m, n)=1$. The fact that $q_i(m, n)$ can be negative indicates that the inverse of the error is not physical and thus can not be implemented by a deterministic quantum operation. To cancel the effect of the error operator, we rewrite the decomposition as follows
\begin{eqnarray}
R_{{\mathcal E}_\phi(m,n)}^{-1}=C_{YY_\varphi(m,n)}\sum_{i=0}^{15}p_i(m,n){\rm sgn}\left[q_i(m,n)\right]R_{P_i},
\end{eqnarray} 
where $p_i(m,n)=C_{YY_\varphi(m,n)}^{-1}\left|q_i(m,n)\right|$ are well-defined probabilities with the cost $C_{YY_\varphi(m,n)}=\sum_{i=0}^{15}\left|q_i(m,n)\right|$, and ${\rm sgn}(\cdot)$ is the sign function.

Then, we can use the Monte-Carlo sampling to compute the PEC error-mitigated results. Here we consider a target quantum circuit consisting of $N_g$ number of two-qubit gates $YY_{\varphi}(m,n)$, with the $g$-th gate labeled as $G_g$, each of which is fully characterized and the inverse-error decomposition is written as $R_{G_g}=\sum_iq_{i,g}R_{P_i}R_{G_g^{\rm ns}}$ with $q_{i,g}$ being the quasi-probability of the two-qubit Pauli operator $P_i$. The full set of the random circuits to be implemented in the experimental device denoted as $\left\{{\mathcal S}_{\mathbf i}:{\mathbf i}=\left(i_1,\ldots, i_{N_g}\right)\right\}$, contains $16^{N_g}$ different circuits, each of which is obtained by adding $P_{i_g}$ right after $G_g$. To obtain the ideal expectation values, each random circuit ${\mathcal S}_{\mathbf i}$ is assigned with a probability $\propto\prod_g\left|q_{i_g, g}\right|$ and a sign ${\rm sgn}_{\mathbf i}\equiv\prod_g{\rm sgn}\left(q_{i_g,g}\right)$. As it is infeasible to implement all these circuits, we use the Monte Carlo sampling to obtain average values over the random circuits. Specifically, we generate $N_s$ random circuits by sampling $P_{i_g}$ from the probability distribution $\propto\left|q_{i_g, g}\right|$ and obtain the expectation value $\langle\mu\rangle_{\mathbf i}$ of an observable $\mu$ for the circuit ${\mathcal S}_{\mathbf i}$ by averaging 300 repetitions. The ideal expectation value $\langle\mu\rangle$ is then calculated by 
\begin{eqnarray}
\langle\mu\rangle=\frac{C}{N_s}\sum_{s=1}^{N_s}{\rm sgn}_{{\mathbf i}_s}\langle\mu\rangle_{{\mathbf i}_s},
\label{eq:9}
\end{eqnarray}where the cost is given by $C\equiv\prod_{g=1}^{N_g} C_g$ with the cost of the $g$-th gate being $C_g\equiv\sum_{i_g=0}^{15}\left|q_{i_g,g}\right|$.

\begin{figure*}[htp!]
\centerline{\includegraphics[width=\textwidth]{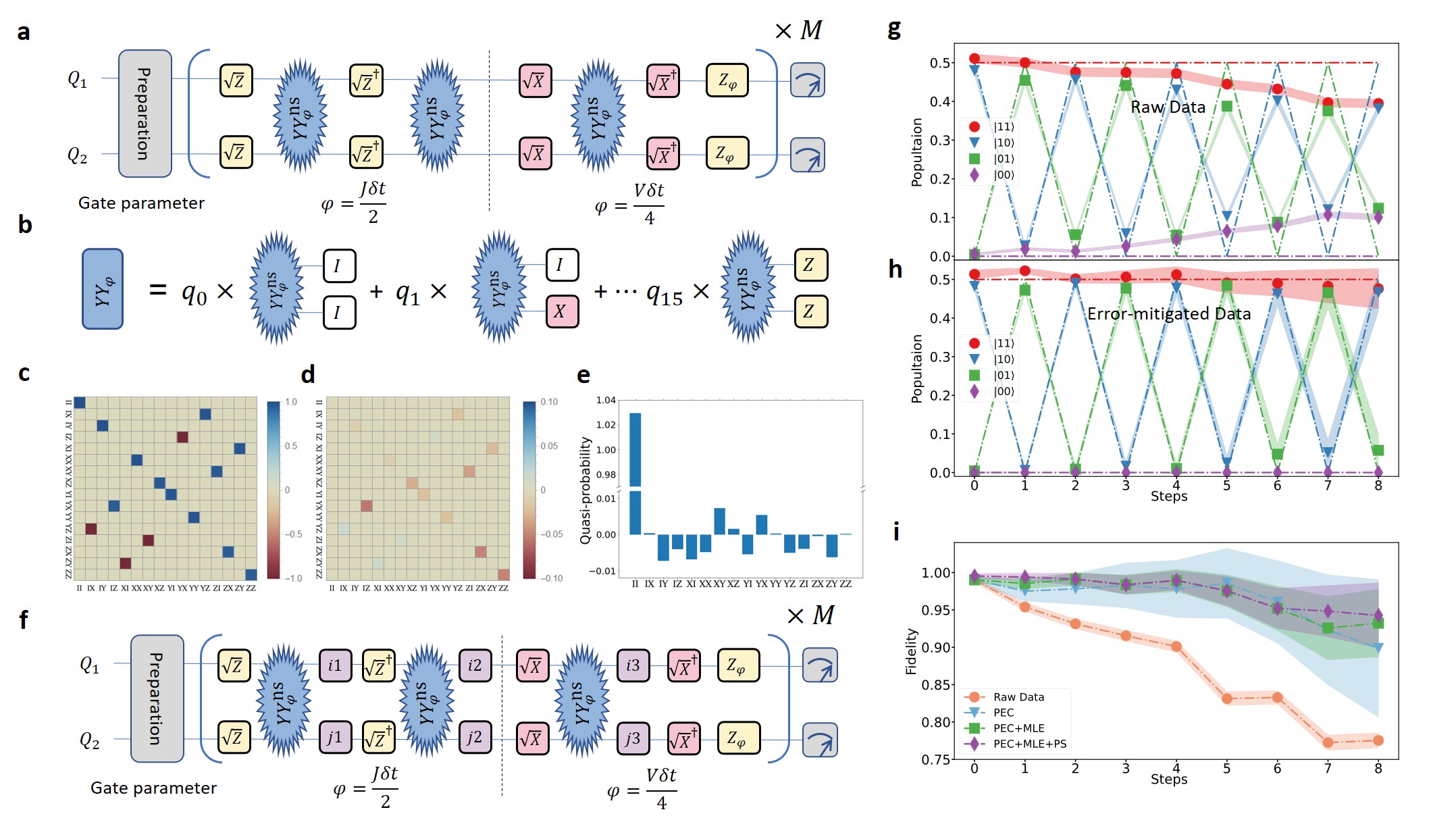}}
\caption{
\textbf{(a)} Quantum circuit of Trotter expansion for the simulation of dynamics of two spinless fermions. Two qubits are initialized to $(\ket{11}+\ket{10})/\sqrt{2}$ after the state preparation, then we apply $M$ times of Trotter steps consisting of three two-qubit gates ${YY}^{\rm ns}$ in each step, which is not perfect. Here, $J$ and $V$ represent the strength of nearest-neighbor tunneling and nearest-neighbor interaction.
\textbf{(b)} Probabilistic error cancellation (PEC) scheme. The imperfections of the gate ${YY}^{\rm ns}_{\varphi}$ can be described by error channels acting on the ideal gate. The ideal gate ${YY}_{\varphi}$ can be realized by including the inverse of the error channels, which can be decomposed into combinations of basis operations with quasi-probabilities. Under the Pauli error assumption, there are 16 basis operations, where the list is shown in \textbf{(c),(d),(e)}. \textbf{(c)} Pauli transfer matrix and \textbf{(d)} Deviation of the experimental ${YY}^{\rm ns}_{\varphi}$ gate, which is characterized through QPT. 
\textbf{(e)} Quasi-probabilities in the decomposition of the inverse error operations of experimental ${YY}^{\rm ns}_{\varphi}$ gate.
\textbf{(f)} Error-mitigated quantum circuits by the PEC. We implement the inverse error operations by sampling with the renormalized probabilities of quasi-probabilities as $p_{i}=|q_{i}|/\sum_{i=0}^{15}|q_{i}|$, where $i$ indicates what number it corresponds to among basis operations. \textbf{(g)} Experimental data and \textbf{(h)} Error-mitigated data for $V=2J$. The solid points represent the state populations measured from the experiment and the error-mitigated data after applying PEC, MLE, and PS methods. The dashed-dot lines represent ideal numerical simulation data with only Trotter errors, where the same colors correspond to the same experimental data points. The shadows around the data points represent error bars.
\textbf{(i)} Population fidelity of experiment data and error-mitigated data. The fidelity is calculated by $|\sum_k \sqrt{P_{k,\textrm{ideal}}P_k}|^2$, which $k$ represents the state basis. The population fidelities after each error-mitigation method are presented. All shadows in (g)-(i) represent the standard deviation of 1000 samples generated from the raw experimental data, using the bootstrapping method.
}
\label{fig:2}
\end{figure*}

\section{Error mitigation and simulation of two spinless fermionic modes}
We now discuss the simulation of two spinless fermionic modes, where the qubit-system Hamiltonian Eq.~(\ref{eq:3}) reduces to 
\begin{eqnarray}
H=\frac{J}{2}\left(\hat\sigma_1^x\hat\sigma_{2}^x+\hat\sigma_{1}^y\hat\sigma_{2}^y\right)+\frac{V}{4}\left(\hat\sigma_1^z\hat\sigma_{2}^z+\hat\sigma_{1}^z+\hat\sigma_{2}^z\right).
\label{eq:Ham2}
\end{eqnarray} With the Trotter-Suzuki expansion of Eq.~(\ref{eq:4}), the evolution of Hamiltonian for two spinless fermionic modes can be experimentally realized with the specific quantum circuit shown in Fig.~\ref{fig:2}a. 
For the singly occupied fermion in one of the sites, the fermion oscillates between two sites caused by the nearest-neighbor tunneling.
For the doubly occupied fermions, there will be no dynamics due to the Pauli exclusion principle of fermions. We apply $M=8$ Trotter steps  and each step consists of three $YY_\varphi$ gates and ten single-qubit rotations. The whole quantum circuit consists of twenty-four two-qubit gates and eighty-one single-qubit gates, where the additional one is for the initial state preparation. We set the nearest-neighbor tunneling and interaction strength as $V=2J$. 

Errors caused by gate imperfection accumulate as the number of Trotter steps increases and the fidelity of evolved dynamics would be degraded. We mitigate the errors and improve the results of dynamics. 
Since it is a two-qubit system, only the one native $YY_{\pi/4}$ gate needs to be mitigated. The ideal $YY_{\pi/4}$ gate can be decomposed as $R_{{YY}_{\pi/4}}=\sum_{i=0}^{15} q_{i} R_{P_i} R_{YY^{\rm ns}_{\pi/4}}$  under Pauli error assumption, which is visualized as Fig.~\ref{fig:2}b.
The experimental $YY^{\rm ns}_{\pi/4}$ gate is characterized by QPT. The experimental results of the PTMs of the $YY^{\rm ns}_{\pi/4}$ gate and the deviation from the ideal one obtained by $R_{YY^{\rm ns}_{\pi/4}}-R_{{YY}_{\pi/4}}$ are shown in Figs.~\ref{fig:2}c and d, where the average gate fidelity of $R_{YY^{\rm ns}_{\pi/4}}$ is $0.9811\pm0.0047$ \cite{nielsen2002simple,greenbaum2015introduction}. Here, we assume the initial state preparation is perfect and apply the detection error correction method \cite{shen2012correcting} for the measurement results, which makes state preparation and measurement (SPAM) errors negligible. The quasi-probabilities, i.e. the coefficients $q_i$ of the two-qubit Pauli operator $P_i$ in the decomposition of $R_{\mathcal E}^{-1}$, are shown in Fig.~\ref{fig:2}e. With the values of the quasi-probabilities $q_i$, we obtain the cost $C_{YY_{\pi/4}}=1.083$ for $R_{YY^{\rm ns}_{\pi/4}}$. 

Then, we generate $N_s=1000$ random circuits by Monte-Carlo sampling illustrated in Fig.~\ref{fig:2}f. With the method described in Eq.~(\ref{eq:9}), we obtain the dynamical evolutions of state populations $P_{\ket{ab}}$ for two spinless fermionic modes, where the observables are $\mu= \ket{ab}\bra{ab} $ with occupied fermion $a(b) = 1$ or no occupied fermion $a(b) = 0$ for each mode. The error-mitigated state populations after PEC can be negative values, which are unphysical. We address this problem by imposing constraints based on general properties of probabilities, i.e. $0\leq P_{\ket{ab}}\leq 1$ and $\sum_{a,b}P_{\ket{ab}}=1$, and obtain physical state populations by the maximum-likelihood estimation (MLE). Specifically, we minimize the norm distance between the physical populations and the PEC error-mitigated ones. On top of PEC and MLE, we also perform post-selection (PS) with respect to the symmetry constraints of the model Hamiltonian, i.e. the conservation law of the number of fermions. 
To demonstrate the effect of each of the error-mitigation techniques, i.e. PEC, MLE, and PS, we apply them step by step and obtain a set of error-mitigated state populations after each step.

The experimental results of fermionic dynamics are shown in Figs.~\ref{fig:2}g and h for the cases without and with error mitigation, respectively. We use the initial two-qubits state of  $(\ket{11}+\ket{10})/\sqrt{2}$, which contains the superposition of two occupied fermions and a single occupied fermion. The initial states simultaneously reveal the time evolutions of both single and double fermions, since no interference occurs between both dynamics. As shown in both Figs.~\ref{fig:2}g and h, the state $\ket{11}$ and $\ket{00}$ do not change and there is an oscillation between the state $\ket{01}$ and $\ket{10}$, which is the hallmark of fermion dynamics \cite{barends2015digital}. We can see that without error mitigation, the contrast between the experimental populations
(solid points) clearly decays compared with the numerical simulation under the Trotter step (hollow points) as shown in Fig.~\ref{fig:2}g, and the error-mitigated results are closer to the ideal numerical simulations as shown in Fig.~\ref{fig:2}h. The population fidelities depending on Trotter steps without and with the three error-mitigation schemes are shown in Fig.~\ref{fig:2}i. The fidelity of the gate without and with error mitigation are $0.9895\pm0.0021$ and $0.9975\pm0.0009$, which are obtained by exponential fitting of data. 

We note that the main improvement in fidelities comes from the PEC method. It is because our characterization of the gate is a good representation of the actual gate. We confirm it by simulating  the dynamics of two spinless modes with the experimentally constructed PTM of $R_{YY^{\rm ns}_{\pi/4}}$. The comparisons between the simulation with the noisy PTM and experimental data clearly validate the
Pauli error assumption and accuracy of measured $R_{YY^{\rm ns}_{\pi/4}}$ (see Appendix B). We also note that MLE reduces the error bar of the results after PEC, which may compensate for the drawback of PEC in the aspect of variance. All the error bars are estimated by bootstrapping method (see Appendix A). 

\begin{figure*}
\centerline{\includegraphics[width=1\textwidth]{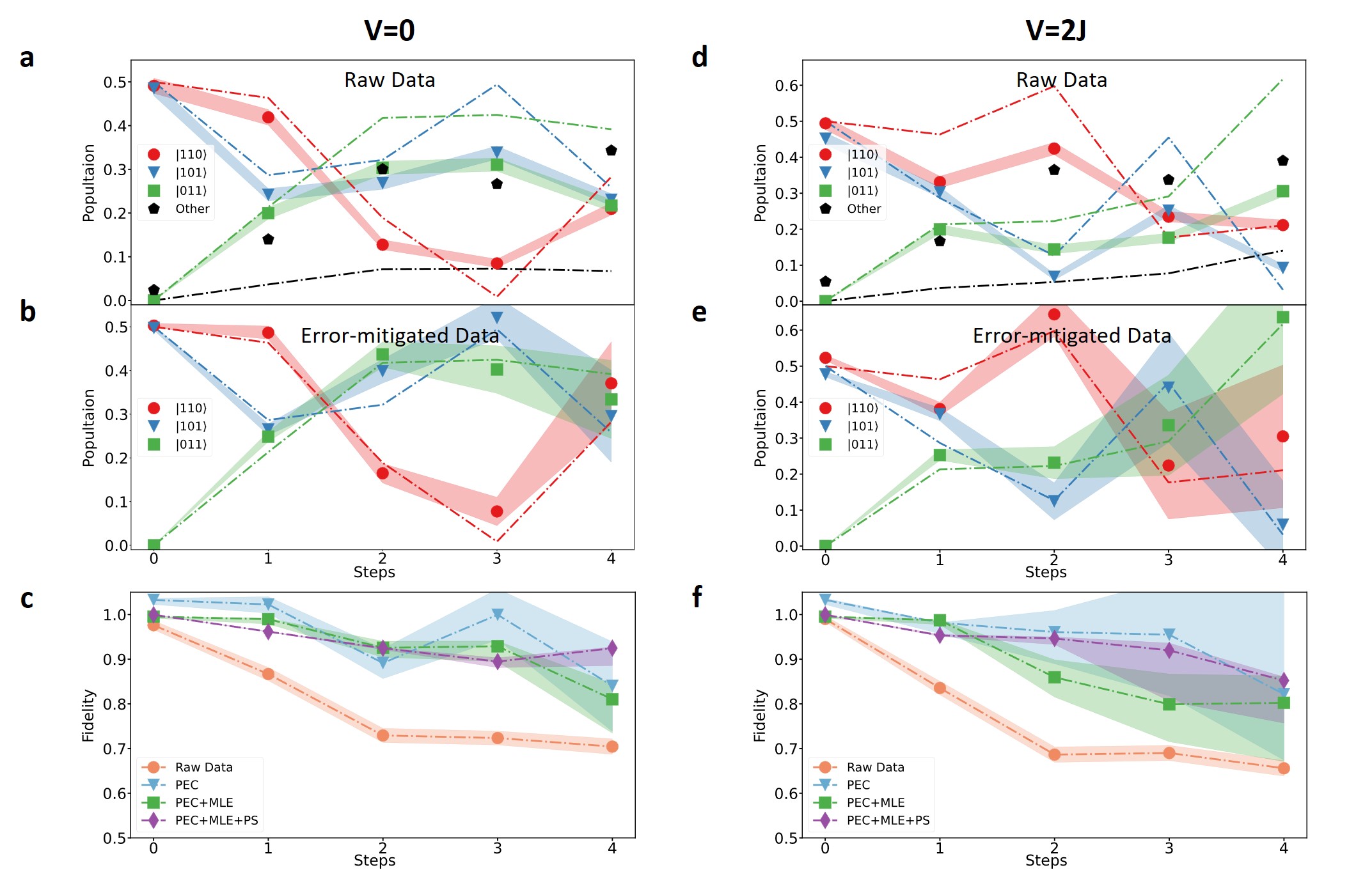}}
\caption{ Dynamics of three spinless fermionic modes with the initial state of
$(\ket{101}+\ket{110})/\sqrt{2}$ \textbf{(a)}-\textbf{(c)} for no interaction $V=0$ and \textbf{(d)}-\textbf{(f)} for $V=2J$. 
\textbf{(a),(d)} Experimentally measured populations without error mitigation and \textbf{(b),(e)} with error mitigation. The solid points represent the experimentally-measured state populations and the dashed-dot lines represent
the numerical simulation with the given Trotter steps. \textbf{(c),(f)} The population fidelities after each error-mitigation method as PEC, MLE, and PS. All shadows in \textbf{(a)-(f)} represent the standard deviation of 1000 samples generated from the raw experimental data, using the bootstrapping method.}
\label{fig:3}
\end{figure*}

\section{simulation of three spinless fermionic modes}
We implement the time evolution of three spinless fermionic modes as shown in Fig.~\ref{fig:device_schematic}a. The quantum circuit in Fig.~\ref{fig:device_schematic}d, obtained by the Trotter-Suzuki expansion, simulates the dynamics of the system. We initialize the three-qubit state to $(\ket{101}+\ket{110})/\sqrt{2}$, which includes two fermions in three sites. The fermions propagate in the system caused by the nearest-neighbor tunneling to other sites. The dynamical evolution would be restricted in the subspace spanned by $\{\ket{110}, \ket{101}, \ket{011}\}$ because of the conservation law of the number of fermions. 

The dynamics can be different depending on the existence of nearest-neighbor interaction different from the previous case with two sites. In the experiment, we choose $V=0$ and $V=2 J$ for the scenarios without and with nearest-neighbor interactions, respectively. In Figs.~\ref{fig:3}a, b, d, and e, dashed-dot lines show the results of ideal numerical simulation. When $V=0$, the initial state $\ket{110}$ mostly evolves to the state $\ket{101}$ because of the Pauli exclusion principle of fermions. On the other hand, the initial state $\ket{101}$ can evolve to both  $\ket{110}$ and $\ket{011}$ states. Therefore, in the beginning, the overall population of the state $\ket{110}$ reduces, and the population of the state $\ket{011}$ increases as shown in Figs.~\ref{fig:3}a and b. When $V=2J $, we can see the population of $\ket{110}$ is not rapidly decreasing and the state $\ket{101}$ is not populated compared to the case of 
$V=0$. It is because the interaction is effectively attractive in our experiment.  

In experiment, we set $\varphi=\frac{J \delta t}{2}=\frac{\pi}{8}$ of the $YY_\varphi$ gates and apply $M=4$ Trotter steps for both scenarios. With nearest-neighbor interactions, each Trotter step consists of six $YY_\varphi$ gates and twenty single-qubit rotations. The whole quantum circuit consists of twenty-five two-qubit gates, where the additional one is for the initial state preparation, and eighty single-qubit gates. Then, we apply the error-mitigation methods to the dynamic evolution of both scenarios. There are two types of ${YY}$ gates on different pairs of qubits, i.e., $YY_{\pi/8}(1,2)$ and $YY_{\pi/8}(2,3)$, both of which require characterization and inverse-error decomposition. In the PEC scheme, both ${YY_{\pi/8}(1,2)}$ and ${YY_{\pi/8}(2,3)}$ gates are individually characterized by QPT and the inverse errors are decomposed independently. The average gate fidelities are $0.9779\pm0.0088$ for $R_{YY^{\rm ns}_{\pi/8}(1,2)}$ and $0.9748\pm0.0085$ for $R_{YY^{\rm ns}_{\pi/8}(2,3)}$   (see Appendix B). 

For the case of $V=0$, the experimental results without and with error mitigation are shown in Figs.~\ref{fig:3}a and b, respectively. We can see the error-mitigated results are more consistent with ideal numerical simulations. The fitted population fidelities per gate without and with error mitigation are $0.9792\pm0.0161$ and $0.9942\pm0.0067$, as shown in Fig.~\ref{fig:3}c. We note that the main improvement comes from the PEC method. This is consistent with the comparison results between the simulation with the noisy PTMs of both $YY$ gates and experimental data, which indicates the measured PTMs properly characterize the two-qubit gate errors (see Appendix C). Similar conclusions are obtained for the scenario with the nearest-neighbor interaction of $V=2J$, as shown in Figs.~\ref{fig:3}d, e and f. The fitted population fidelities per gate without and with error mitigation are $0.9827\pm0.0142$ and $0.9889\pm0.0080$. The two-qubit gate errors in the three-qubit system are larger than that in the two-qubit system, which leads to the larger cost as $C_{YY_{\pi/8}(1,2)}=1.157$ and $C_{YY_{\pi/8}(2,3)}=1.171$. Therefore, we sample more random circuits as $N_s=1500$ to reduce the error bars of the final error-mitigated results. 

We note that the fitted population fidelities after PEC could be unphysical because of the possible negative quasi-probabilities and the cost being over 1. As shown in Figs.~\ref{fig:3}c and f,  fidelities greater than 1 after applying PEC exist at the beginning of the three-fermion experiments. With the help of the MLE method based on the general properties of probabilities, we can make the population fidelities physical and reasonable, which is, below 1 as shown in green dots of Figs.~\ref{fig:3}c and f. In addition, MLE reduces the error bars of the results after PEC, which may compensate for the drawback of PEC in the aspect of variance. After applying PEC, the standard deviation of the fourth Trotter step become unexpectedly large. However, the MLE method suppresses the standard deviation by a few factors in this case. 

\begin{figure*}
\centerline{\includegraphics[width=1\textwidth]{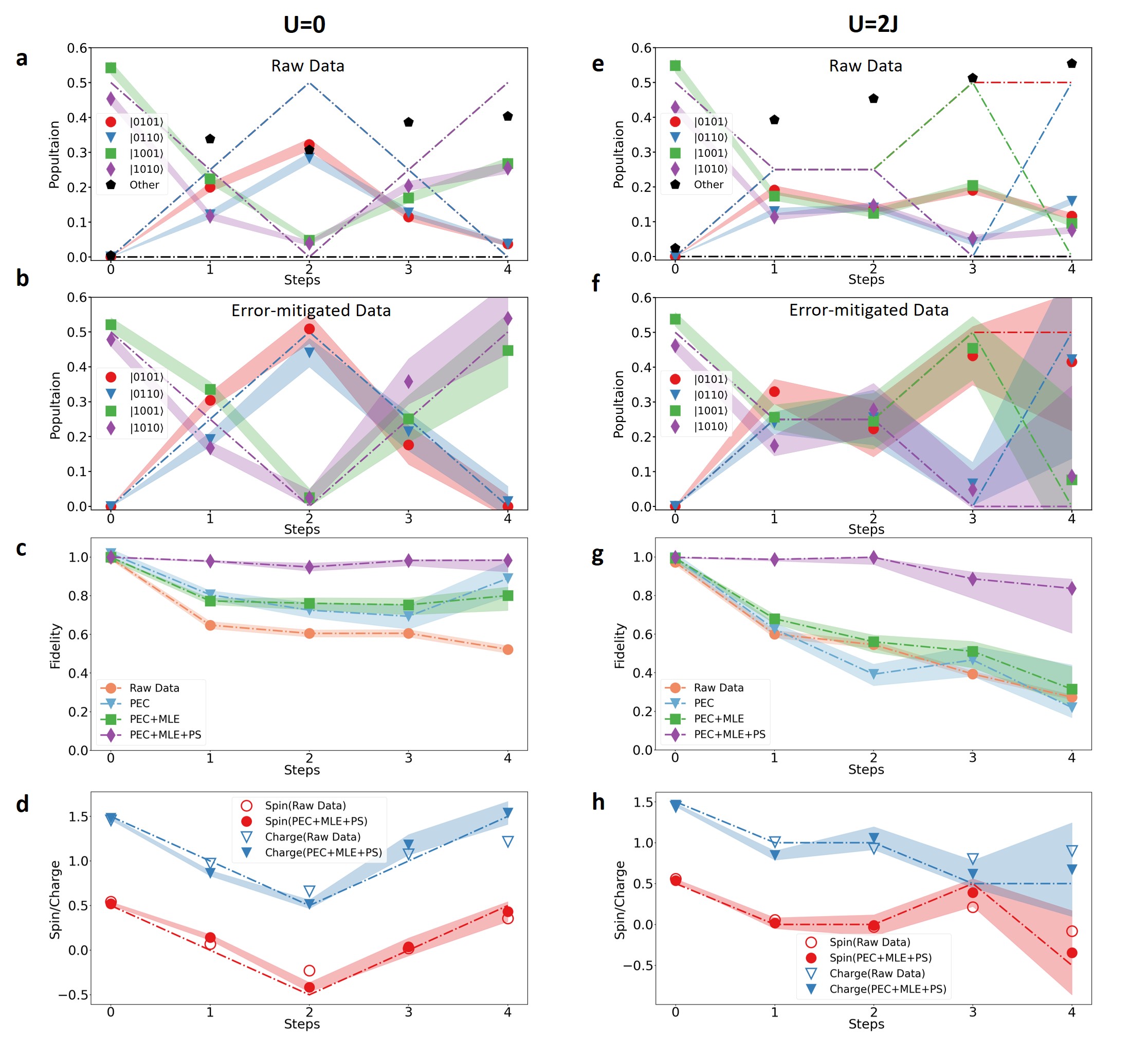}}
\caption{ Dynamics of four fermionic modes mapped by two fermion sites and spins with the initial state of
$(\ket{1001}+\ket{1010})/\sqrt{2}$ \textbf{(a)}-\textbf{(d)} for no interaction $V=0$ and \textbf{(e)}-\textbf{(h)} for $V=2J$. 
\textbf{(a),(e)} 
Experimentally measured population without error mitigations and \textbf{(b),(f)} with error mitigation. The solid points represent the experimentally-measured state populations and the dashed-dot lines represent
\textbf{(c),(g)} The population fidelities after each error-mitigation method as PEC, MLE, and PS. \textbf{(d),(h)} Fermionic spin and charge population. The hollow and filled symbols represent the experimental data without and with error mitigations, respectively. The dashed-dot lines represent the ideal numerical simulation with Trotter errors. All shadows represent the standard deviation of 1000 samples generated from the raw experimental data, using the bootstrapping method. 
}
\label{fig:4}
\end{figure*}

\section{Simulation of four fermionic modes} 
We implement the simulation of four fermionic modes with spins by encoding qubits $Q_1Q_3$ and $Q_2Q_4$ as two fermionic sites, where each site can contain two fermionic modes with different spins, as shown in Fig.~\ref{fig:device_schematic}b. We initialize the four qubits state to $(\ket{1001}+\ket{1010})/\sqrt{2}$, which is corresponding to the superposition state between $\ket{\uparrow}_1 \ket{\downarrow}_2 $ and $\ket{\uparrow}_1 \ket{\downarrow}_1$ as shown in Fig.~\ref{fig:device_schematic}b. The two fermions interact and exchange on the four modes caused by the nearest-neighbor tunneling and on-site interaction on subspace states $\{\ket{0101},\ket{0110},\ket{1001},\ket{1010}\}$ with conserved fermions and spins. The corresponding quantum circuit is illustrated in Fig.~\ref{fig:device_schematic}e. Here, we expect the different dynamic behaviors of spin and charge depending on the on-site interaction strength. Spin and charge of the fermions are defined as ${\rm spin}= n_{j,\uparrow}-n_{j,\downarrow}$ and $ {\rm charge}= n_{j,\uparrow}+n_{j,\downarrow}$ with $n_{j,\uparrow(\downarrow)}$ being the fermion number of spin-up (down) on the $j$-th site. 

To demonstrate the different behavior of spin and charge, we consider two scenarios $U=0$ and $U=2J$, which are related without and with on-site interactions, respectively. For the case of no on-site interaction, the spin and the charge of fermions should show the same dynamics. But if there is on-site interaction, the dynamics for spin and charge should show a different behavior, which is related to spin-charge separation for low energy excitation \cite{tarruell2018quantum,vijayan2020time,arute2020observation,senaratne2022spin}. For the case of no on-site interaction, $U=0$, the dynamics of fermions with spins are similar to those without spins, which are discussed in Figs.~\ref{fig:2}g and h. The state $\ket{\uparrow}_1 \ket{\downarrow}_2$($\ket{1001}$) is going back and forth with the other state $\ket{\uparrow}_2 \ket{\downarrow}_1$($\ket{0110}$). Likewise, the state $\ket{\uparrow}_1 \ket{\downarrow}_1$($\ket{1010}$) is oscillating with the other state $\ket{\uparrow}_2 \ket{\downarrow}_2$($\ket{0101}$). This leads to equal amounts of charge and spin change at each site. However, when there exists on-site interaction, the dynamics of those fermionic states are no longer simple oscillations, and the state $\ket{\uparrow}_1 \ket{\downarrow}_2$, which is not affected by the on-site interaction, and the state $\ket{\uparrow}_1 \ket{\downarrow}_1$, which is strongly influenced by the on-site interaction, show completely different dynamics.

In experiment, we set $\varphi=\frac{J \delta t}{2}=\frac{\pi}{8}$ and apply $M=4$ Trotter steps for both scenarios. With on-site interactions, each step consists of six $YY_\varphi$ gates and twenty single-qubit rotations. The whole quantum circuit consists of twenty-five two-qubit gates, where the additional one is for the initial state preparation, and eighty single-qubit gates. In order to apply error mitigations, in particular, the PEC scheme, we characterize four types of $YY$ gates on different pairs of qubits, i.e., $YY_{\pi/8}(1,2)$, $YY_{\pi/8}(3,4)$, $YY_{\pi/8}(1,3)$, and $YY_{\pi/8}(2,4)$, individually and decompose their inverse errors independently. The average gate fidelities are $0.9755\pm0.0096$  for $R_{YY^{\rm ns}_{\pi/8}(1,2)}$, $0.9706 \pm0.0085$ for $R_{YY^{\rm ns}_{\pi/8}(3,4)}$, $0.9720\pm0.0092$ for $R_{YY^{\rm ns}_{\pi/8}(1,3)}$ and $0.9744\pm0.0091$ for $R_{YY^{\rm ns}_{\pi/8}(2,4)}$ (see Appendix B). 

For the case of $U=0$, the experimental results without and with error mitigation are shown in Figs.~\ref{fig:4}a and b, respectively.  We can see the error-mitigated results are more consistent with the numerical simulation. The fitted population fidelities per gate without and with error mitigation are $0.9658\pm0.0426$ and $0.9993\pm0.0053$, respectively, as shown in Fig.~\ref{fig:4}c. In Fig.~\ref{fig:4}d, the same behaviors of spin and charge, where the amounts of charge and spin changes are the same in the dynamics, are clearly shown with error mitigation, where the net charge and the net spin at site 1 are displayed. We can observe similar improvements in the dynamics with on-site ($U=2J$) interaction, as shown in Figs.~\ref{fig:4}e, and f. In Fig.~\ref{fig:4}g, the fitted population fidelities per gate without and with error mitigation are $0.9551\pm0.0285$ and $0.9925\pm0.0081$, respectively. With on-site interaction, the different behaviors of spin and charge are more clearly shown after error mitigations as shown in Fig.~\ref{fig:4}h.

We note that the PEC scheme alone does not recover the dynamics for the four fermionic modes as shown in Figs.~\ref{fig:4}c and g different from the simulations with two and three fermionic modes. The main improvements in fidelities come from the post-selection methods based on the fermion-number-conservation assumption. One reason is that the experimental PTMs of two-qubit gates in the four-qubit system do not properly characterize the imperfections of those gates. It can be seen from the inconsistency of the numerically simulated data with the PTMs of two-qubit gates and experimental results (see Appendix C). The PTMs in our experiments cannot accommodate infidelities from cross-talks between qubits and time-correlated errors, that is, the performance of the gates during characterization can be different from those of actual gates in the quantum simulation. We note that coherent errors beyond the Pauli-error model assumption also are not captured by our experimental PTMs, which partially characterize errors related to Pauli errors. However, errors from cross-talks and time-correlations would be more important factors because coherent errors have no obvious reason to be seriously worse with system size. We also note that the number of sampling is greatly increased for applying the PEC  method in the four-qubit system. With the measured fidelities, the costs increase to $C_{YY_{\pi/8}(1,2)}=1.211$, $C_{YY_{\pi/8}(3,4)}=1.228$, $C_{YY_{\pi/8}(1,3)}=1.228$ and $C_{YY_{\pi/8}(2,4)}=1.223$. The experimental Monte-Carlo sampling as $N_s=2000$ is not sufficient, which is limited by the time-consuming running of massive random circuits in physical hardware.%

\begin{figure*}[htp!]
\centerline{\includegraphics[width=1\textwidth]{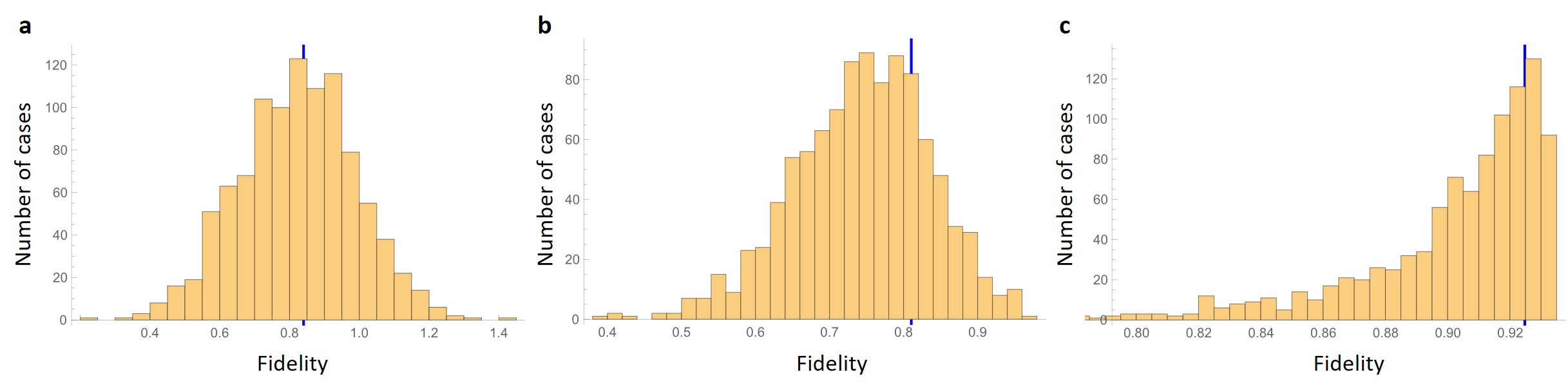}}
\caption{Fidelity distribution at the fourth Trotter step in the simulation of three spinless fermions of all the samples generated from the bootstrap method after \textbf{(a)} PEC method, \textbf{(b)} MLE method, and \textbf{(c)} PS method. The black lines in all the figures represent the original calculated results. Here each figure contains 1000 samples.
}
\label{fig:S1}
\end{figure*}

\begin{figure}[htp!]
\centerline{\includegraphics[width=0.5\textwidth]{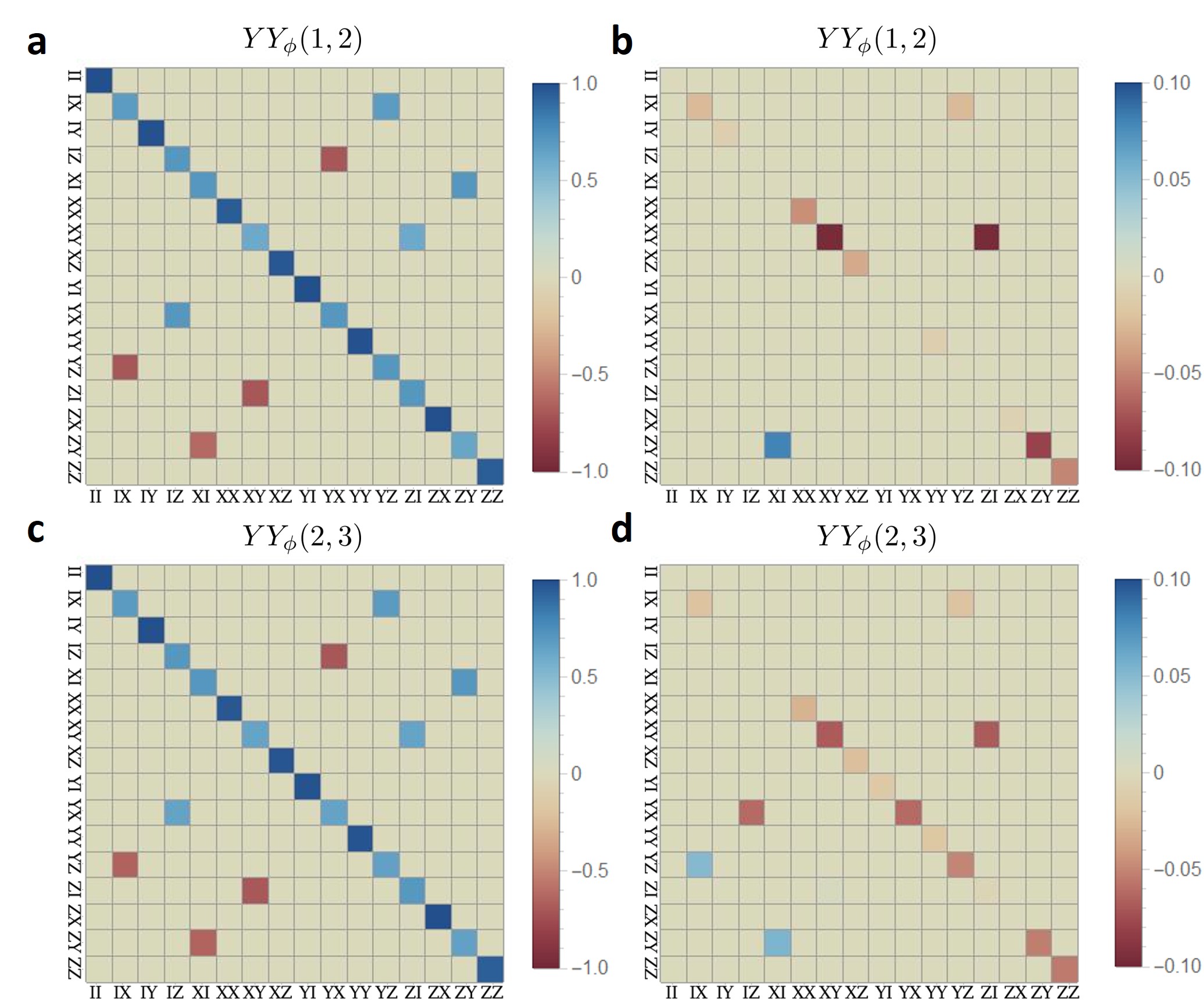}}
\caption{PTMs and the deviation from the ideal $YY_{\pi/8}$ gate for \textbf{(a)}-\textbf{(b)} ${YY}_{\varphi}(1,2)$ and \textbf{(c)}-\textbf{(d)} ${YY}_{\varphi}(2,3)$ in the simulation of three spinless fermions.  
}
\label{fig:S2}
\end{figure}

\section{Discussion and Outlook}
We have applied the PEC method to improve the simulation of interacting fermions on up to four trapped ion qubits. We have observed the improvement of the simulation close to an order of magnitude in fidelities. With four qubits, we are able to observe the different dynamics of spin and charges for the interacting fermions.

Although PEC is the fundamental method in the experiment, several other error migration methods have been developed and applied. One of them is using the positive probability constraint. One intrinsic problem in the PEC method is that the probabilities of quantum states can be negative and not necessarily normalized. We impose constraints such that the probabilities of fermion states in the dynamics are positive and normalized, and determine the probabilities through the maximum-likelihood method. This method guarantees that observables are physically reasonable and can suppress statistical errors. 

We also apply the symmetry constraint such that the total number of fermions and spins are conserved. In the simulation with four qubits, the majority of infidelity comes from the populations outside the number-conserving states. The constraints of total fermions conservation resolve the problem of population leakages. However, it is questionable if the number-conservation constraints would be scalable since the leakage can increase with the system and circuit depth. 

From our experimental demonstration, although PEC incorporating constraints recoveries the dynamics of four trapped-ion qubits, it is clear that we need further improvements on the method for large-scale implementation. In particular, the drift of laser parameters and spectral crowding causes correlations resulting in remaining errors after PEC. As demonstrated in this work, combining with constraint methods can ease this issue. Recently, several proposals and experiments have been reported to tackle correlations such as using sparse Pauli-Lindblad models \cite{berg2022probabilistic}, learning-based error-mitigation \cite{strikis2021learning}, matrix product operator representation \cite{guo2022quantum}, etc. With all these techniques combined together, NISQ computers can potentially achieve the accuracy required for gaining the quantum advantage in practical applications, such as model inference for nuclear magnetic resonance (NMR) spectroscopy \cite{sels2020quantum} and large-depth QAOA \cite{farhi2014quantum}.

\section{acknowledgments}
This work was supported by the Innovation Program for Quantum Science and Technology under Grants No. 2021ZD0301602, and the National Natural Science Foundation of China under Grants No. 92065205, No. 11974200, No. 12225507, No. 12088101, and No. 12204535.

\begin{figure*}[htp!]
\centerline{\includegraphics[width=1\textwidth]{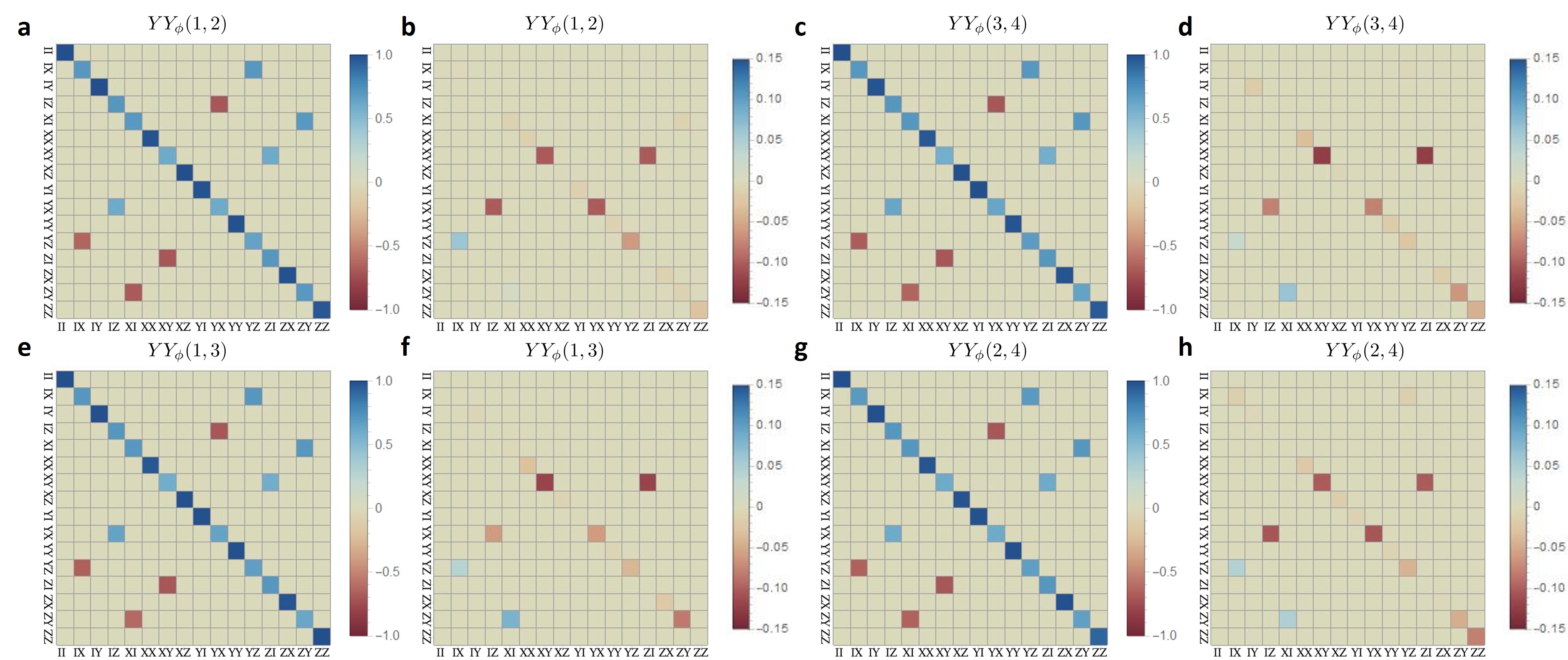}}
\caption{PTMs and the deviation from the ideal $YY_{\pi/8}$ gate for \textbf{(a)}-\textbf{(b)} ${YY}_{\varphi}(1,2)$, \textbf{(c)}-\textbf{(d)} ${YY}_{\varphi}(3,4)$, \textbf{(e)}-\textbf{(f)} ${YY}_{\varphi}(1,3)$ and \textbf{(g)}-\textbf{(h)} ${YY}_{\varphi}(2,4)$ in the simulation of  two fermions with spins.
}
\label{fig:S3}
\end{figure*}

\begin{figure*}[htp!]
\centerline{\includegraphics[width=1\textwidth]{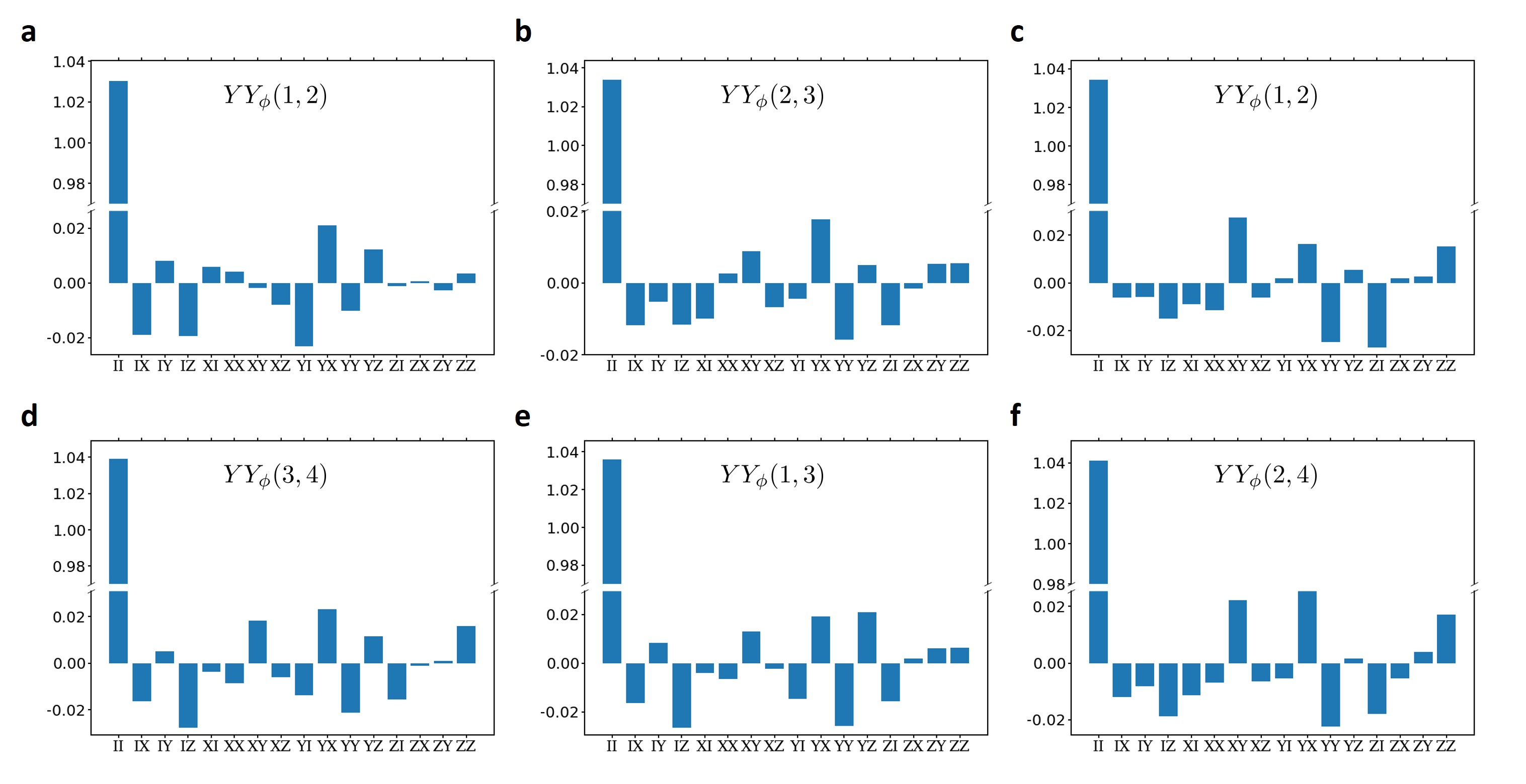}}
\caption{
Quasi-probabilities in the decomposition of the inverse error operations of experimental ${YY}_{\pi/8}$ gate. \textbf{(a)}-\textbf{(b)} Gate ${YY}_{\varphi}(1,2)$ and ${YY}_{\varphi}(2,3)$ for three spinless fermions simulation. \textbf{(c)}-\textbf{(f)} Gate ${YY}_{\varphi}(1,2)$, ${YY}_{\varphi}(3,4)$, ${YY}_{\varphi}(1,3)$ and ${YY}_{\varphi}(2,4)$ for the simulation of two fermions with spins.
}
\label{fig:S4}
\end{figure*}

\section{Appendix A: Estimating error bars with the bootstrapping method}
We statistically estimated the uncertainty of experimental data with and without error mitigations by using the bootstrapping method. In order to estimate the variance of experimental data, we repeatedly and randomly take out samples from our original data and apply PEC, MLE, and PS methods step by step to mitigate the population errors of each sample. In our work, the size of the samples is the same as the original data sets, and the total number of samples for each data point is usually chosen as 1000. Then we calculate the standard deviation of the state population and the population fidelities after each step with the sets of samples. In order to avoid the appearance of unphysical fidelities (larger than 100\% or less than 0\% after MLE and PS methods), we separated the sets of samples into two parts according to the fidelities of the original data set and calculated the standard deviation of each side, which resulted in unsymmetrical error bars. Fig.~\ref{fig:S1} shows the fidelity distribution (step 4 for the simulation of three spinless fermions, V=0) after each error-mitigation step, which indicates an increasing deviation from the normal distribution after more error-mitigation steps are applied.

\begin{figure}[htp!]
\centerline{\includegraphics[width=0.5\textwidth]{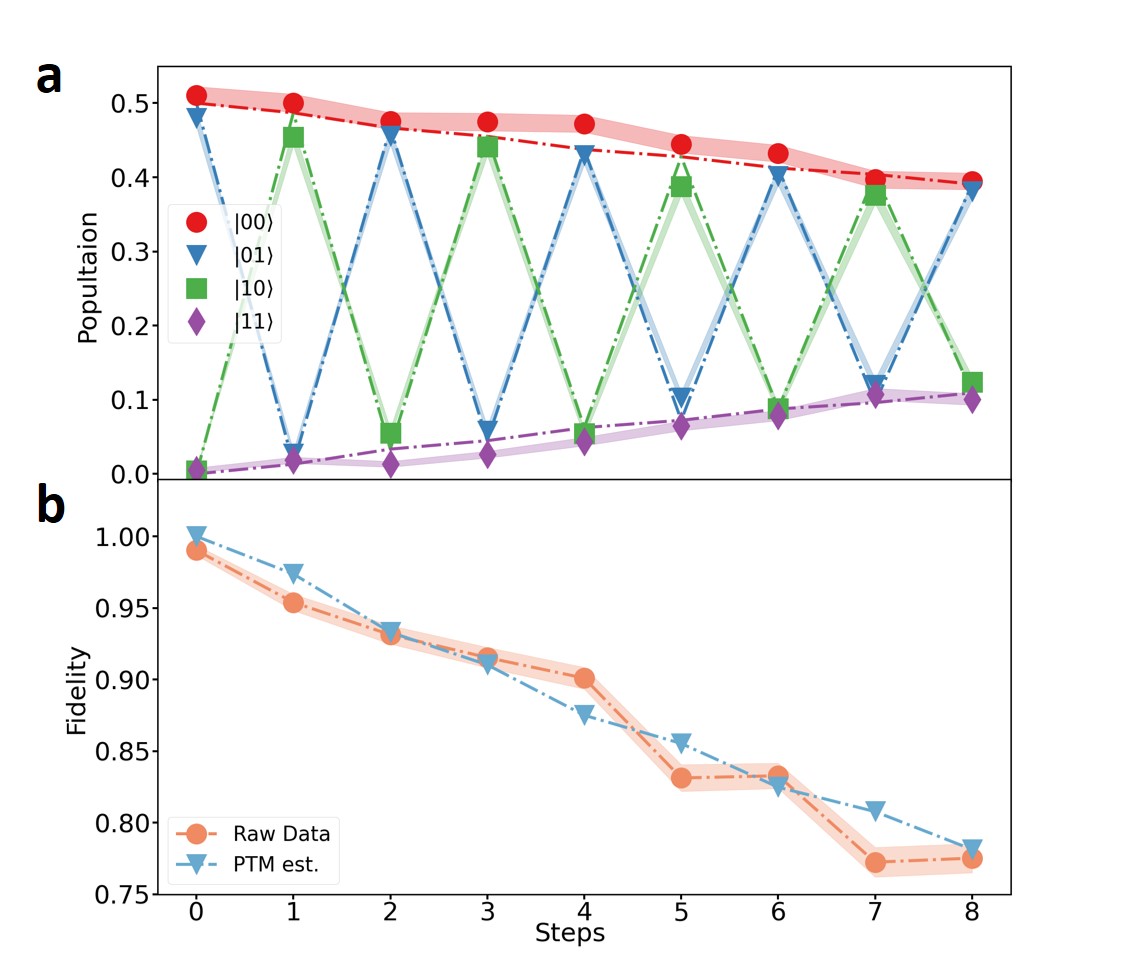}}
\caption{Experimental raw data compared with error-included simulation for two spinless fermions. We simulated the dynamics with noisy gates measured from QPT. \textbf{(a)} Populations of experimental raw data and simulation. The solid points represent the state populations measured from the experiment, and the dashed-dot lines represent numerical simulation data with noisy gates. \textbf{(b)} Population fidelity of experimental raw data and simulation data shown in (a). All the shadows in (a)-(b) represent the standard deviation of 1000 samples generated from the raw experimental data, using the bootstrapping method.
}
\label{fig:S5}
\end{figure}

\section{Appendix B: The PTMs and quasi-probabilities for three and four qubit system} 
The quantum circuits for the simulation of the three spinless fermionic modes (three-site single-component) and four fermionic modes (two-site double-component Fermi-Hubbard model) are shown in Fig.~\ref{fig:device_schematic} (d) and (e). Here up to six two-qubit M{\o}lmer-S{\o}rensen gates $YY_\varphi$ are applied in each Trotter step. We classify the gates depending on related qubit pairs and benchmark each type of gate with the QPT  method. Under Pauli error assumption, we measure 15 different components in $R_{{YY^{\rm ns}}_{\varphi}}$ to get the PTMs of each type of gate as shown in Fig.~\ref{fig:S2} and Fig.~\ref{fig:S3}. The quasi-probabilities for ${YY_{\varphi}}$ gates in the simulation of three spinless fermions and two fermions with spins are shown in Fig.~\ref{fig:S4}, which decides the probabilities of choosing Pauli operations in the Monte Carlo method.

\section{Appendix C: The PTM simulation to understand experimental error models.} 
\begin{figure*}[htp!]
\centerline{\includegraphics[width=1\textwidth]{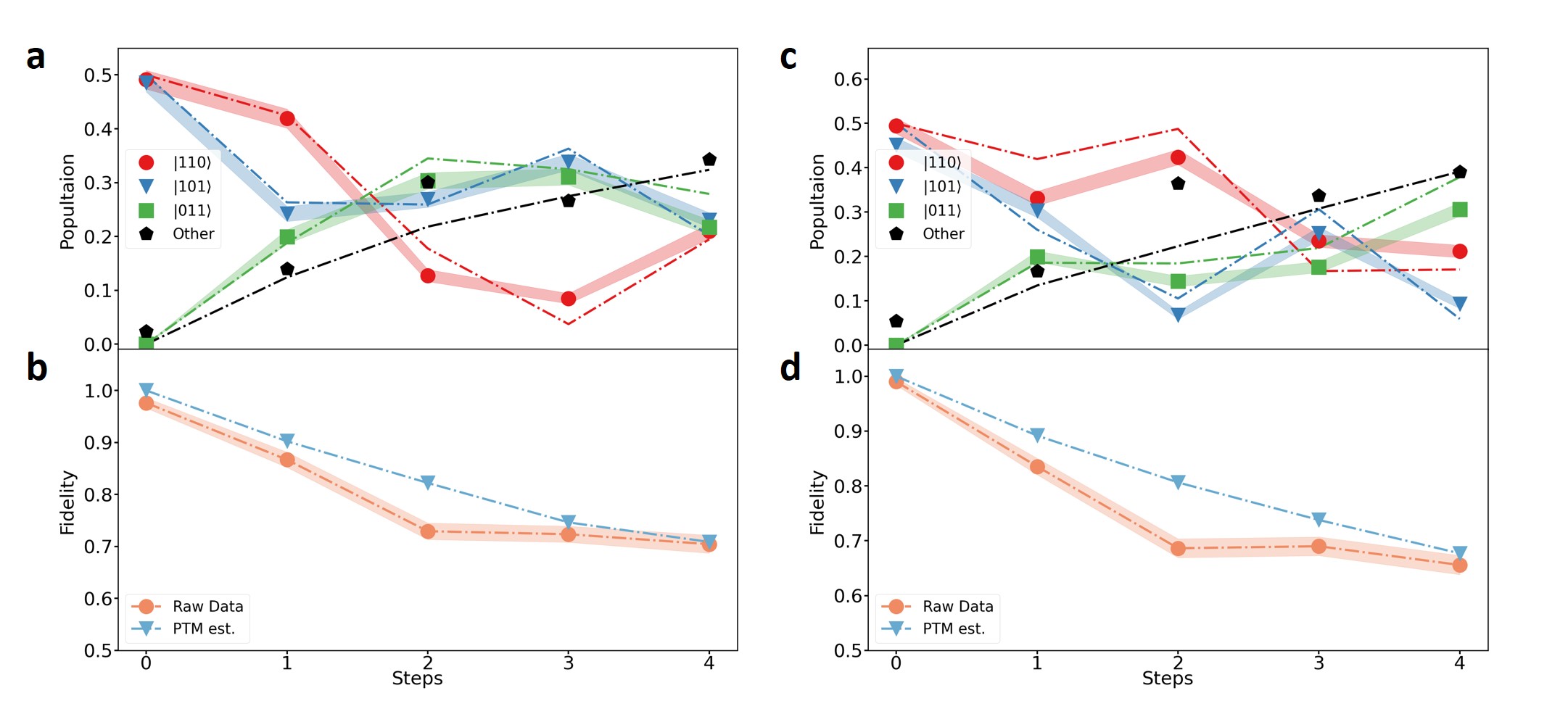}}
\caption{Experimental raw data compared with error-included simulation for three spinless fermions. We simulated the dynamics of $V=0$ case with noisy gates measured from QPT. \textbf{(a)} Populations of experimental raw data and simulation. The solid points represent the state populations measured from the experiment, and the dashed-dot lines represent numerical simulation data with noisy gates. \textbf{(b)} Population fidelity of experimental raw data and simultion data shown in (a). \textbf{(c)}-\textbf{(d)} population and fidelity for $V=2J$ case, the representation is the same as (a)-(b). All the shadows in (a)-(d) represent the standard deviation of 1000 samples generated from the raw experimental data, using the bootstrapping method.
}
\label{fig:S6}
\end{figure*}

\begin{figure*}
\centerline{\includegraphics[width=1\textwidth]{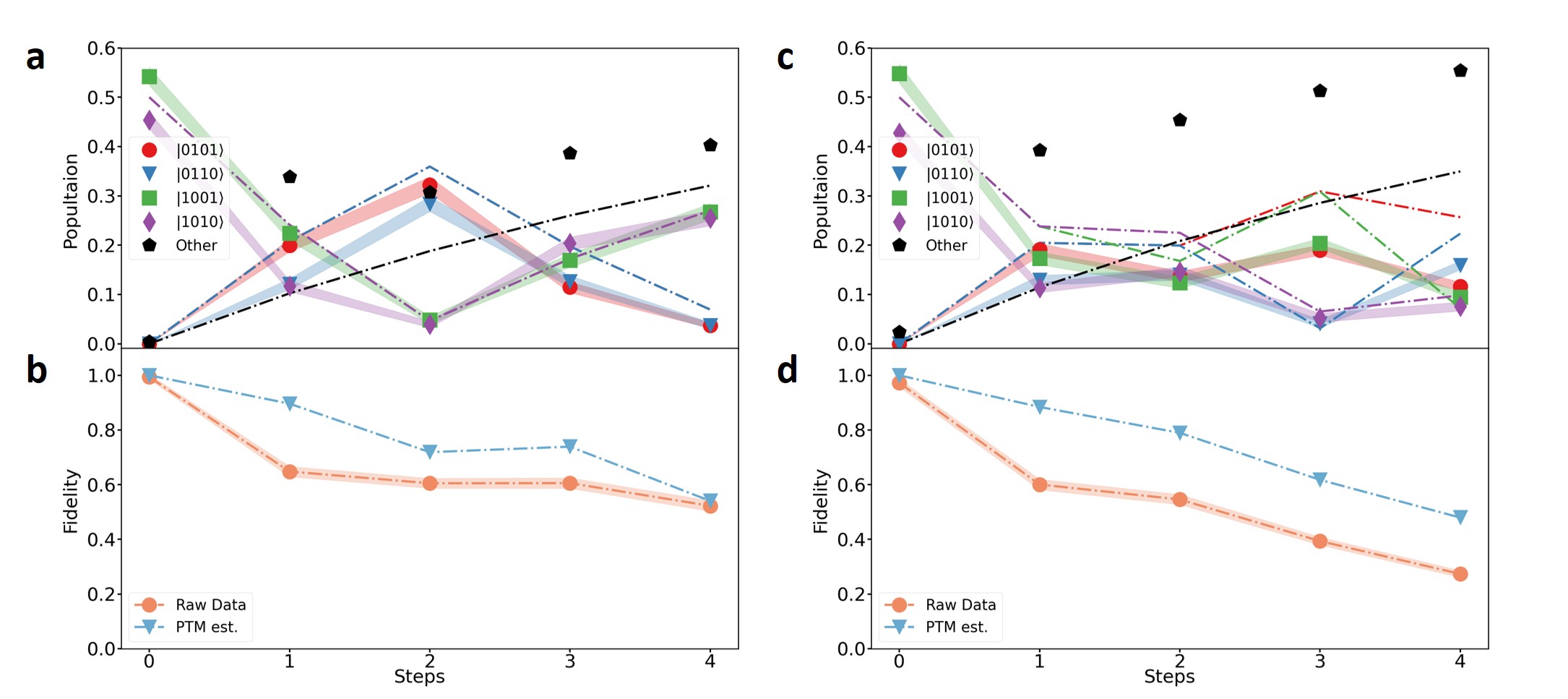}}
\caption{Experimental raw data compared with error-included simulation for two fermions with spins. We simulated the dynamics of $U=2J$ case with noisy gates measured from QPT. \textbf{(a)} Populations of experimental raw data and simulation. The solid points represent the state populations measured from the experiment, and the dashed-dot lines represent numerical simulation data with noisy gates. \textbf{(b)} Population fidelity of experimental raw data and simultion data shown in (a). \textbf{(c)}-\textbf{(d)} population and fidelity for $U=2J$ case, the representation is the same as (a)-(b). All the shadows in (a)-(d) represent the standard deviation of 1000 samples generated from the raw experimental data, using the bootstrapping method.
}
\label{fig:S7}
\end{figure*}

To characterize our noisy two-qubit gates, a partial QPT with Pauli error assumption is applied with 15 measurements out of 144. To verify the assumption that the errors of our gates are consistent with the increasing of Trotter steps, we perform the numerical simulation of the extended Fermi-Hubbard model with the PTMs of the noisy gates. The simulation results are shown in Figs.~\ref{fig:S5}-~\ref{fig:S7} for two spinless fermions, three spinless fermions, and two fermions with spins, respectively. For the simulation of two spinless fermions, with no additional leakage to other states out of the selected subspace, we can get a consistent simulation result with the experimental data, which further convinced the significant improvement of fidelity with the PEC method as shown in the introduction of Trotter errors Fig.~\ref{fig:2}(i). For the simulation results for three spinless fermions and two fermions with spins, the population in other states out of the ideal subspace is introduced because of the Trotter errors. The leakage will be further increased by the noisy gates in the experiment, which needs to be mitigated by MLE and PS methods. As a result, the simulation results shown in Fig.~\ref{fig:S6} and ~\ref{fig:S7} have obvious deviations from the experimental data, which can be considered as a result of the unexpected errors of gates such as crosstalk error.

\strut\newpage

\strut%

\end{document}